\documentclass[preprint,aps]{revtex4}
\usepackage{graphicx}
\usepackage{dcolumn}
\usepackage{bm}

\begin{document}


\title{Proximity effect of vanadium on spin-density-wave magnetism in Cr films\\}

\author{E. Kravtsov}
 \email{evgeny.kravtsov@ruhr-uni-bochum.de}
\affiliation{%
Institut f\"{u}r Experimentalphysik/Festk\"{o}rperphysik, Ruhr-Universit\"{a}t Bochum, D-44780 Bochum, Germany \\
and\\ Institute of Metal Physics, 620219 Ekaterinburg, Russia\\}%
\author{A. Nefedov}%
\author{F. Radu}%
\author{A. Remhof}%
\author{H. Zabel}%
\affiliation{%
Institut f\"{u}r Experimentalphysik/Festk\"{o}rperphysik, Ruhr-Universit\"{a}t Bochum, D-44780 Bochum, Germany \\
}%
\author{R. Brucas}%
\author{B. Hj\"{o}rvarsson}
\affiliation{
Department of Physics, Uppsala University, Box 530, S-75121 Uppsala, Sweden\\
}%
\author{A. Hoser}
\affiliation{ Institut f\"{u}r Kristallographie, RWTH-Aachen,
D-52056 Aachen, Germany\\and\\Institut f\"{u}r
Festk\"{o}rperforschung, Forschungszentrum J\"{u}lich, D-52425
J\"{u}lich, Germany\\
}%
\author{S. B. Wilkins}
\affiliation{ European Commission, JRC, Institute for Transuranium
Elements,  D-76125 Karlsruhe, Germany\\and\\European Synchrotron
Radiation Facility, F-38000
Grenoble, France\\
}%

\date{\today}

\begin{abstract}
The spin-density wave (SDW) state in thin chromium films is well
known to be strongly affected by proximity effects from
neighboring layers. To date the main attention has been given to
effects arising from exchange interactions at interfaces. In the
present work we report on combined neutron and synchrotron
scattering studies of proximity effects in Cr/V films where the
boundary condition is due to the hybridization of Cr with
paramagnetic V at the interface. We find that the V/Cr interface
has a strong and long-range effect on the polarization, period,
and the N\'{e}el temperature of the SDW in rather thick Cr films.
This unusually strong effect is unexpected and not predicted by
theory.
\end{abstract}

\maketitle
\section{\label{sec:level1}Introduction}
The spin density wave (SDW) magnetism in Cr has attracted much
attention since its first discovery in 1958
\cite{{corliss59},{bykov58}}. Its bulk behavior is well
established and  considerable progress has  also been made towards
understanding of the SDW magnetism in alloys with other transition
metals (see reviews \cite{{fawcett88},{fawcett94},{LB86},{LB97}}
and references therein). During the last decade this field has
regained much momentum because of a shift of the main focus
towards thin Cr films and multilayers. The SDW behavior in these
systems was found to display new features, differing drastically
from the bulk ones due to dimensional and proximity effects.

Bulk Cr is an itinerant antiferromagnet displaying an
incommensurate SDW (ISDW) below the N\'{e}el temperature
$T_{N}$=311 K. The SDW consists of a sinusoidal modulation of the
amplitude of the antiferromagnetically ordered magnetic moments.
The SDW is a linear wave, propagating always along \{100\}
directions in the bcc Cr lattice, while the wave-vector is
incommensurate with the Cr lattice periodicity. At the spin-flip
transition temperature T$_{SF}$=123 K, the SDW polarization
changes from longitudinal (LSDW) (T $\leq$ T$_{SF}$, magnetic
moments aligned parallel to the wave propagation direction) to
transverse (TSDW) (T $\geq$ T$_{SF}$, magnetic moments aligned
perpendicular to the wave propagation direction). Elastic strains
and chemical impurities in Cr may cause the SDW to become
commensurate (CSDW). In some cases the CSDW phase may be stable at
temperatures much higher than 311 K
\cite{{fawcett88},{fawcett94}}. Generally, in bulk Cr the SDW
exists in a polydomain state below the N\'{e}el temperature, with
magnetic domains populating all possible \{100\} directions.

Investigations of the SDW in thin Cr films were initially
motivated by the significant interest in exchange coupled Fe/Cr
multilayers displaying a giant magnetoresistance effect
\cite{baibich88}. Today, an extensive literature has evolved,
which traces different aspects of the Cr magnetism in Fe/Cr
systems \cite{Unguris92, Fullerton95, Schreyer95, bodeker98}. From
these studies it was concluded that the formation of SDW magnetism
in thin Cr films is governed by a strong exchange coupling acting
at the Cr/Fe interface on the one hand and by the interface
structure/disorder on the other hand \cite{{pierce99}, {zabel99},
{fishman01}}. Similar observations have also been made for the SDW
behavior in other Cr/ferromagnet systems, such as Cr/Ni and Cr/Co
\cite{bodeker99}.

Although a complete understanding of the SDW magnetism in
Cr/ferromagnetic systems has not been reached yet, attention has,
in part, shifted to layered heterostructures of Cr in contact with
nonmagnetic metals. Among others, the SDW state in Cr/Ag
\cite{meersschaut}, Cr/Cu\cite{bodeker99}, and
Cr/V\cite{almokhtar00, mibu01} has been investigated recently. In
particular, the Cr/V system promises to provide new physics via
the hybridization between very similar Fermi surfaces and the
tunability of the Fermi surface with hydrogen uptake in the
vanadium host lattice.

It is well known that the SDW behavior is connected with the
peculiar features of the Cr Fermi surface providing nesting
vectors between electron and hole sheets of similar shape
\cite{overhauser62}. Doping with V decreases the electron
concentration, and, hence, the magnetic moment of Cr. This is
opposite to doping with Mn atoms, which enhances the magnetic
moment and drives Cr towards CSDW order \cite{fullerton03}.
Comprehensive investigations of bulk CrV alloys
\cite{{hamaguchi65}, {koehler66}, {trego68}} have revealed a
linear decrease of the N\'{e}el temperature with the V
concentration, followed by a corresponding decrease in the SDW
period. The effect was found to be so strong that a concentration
of only 4 at. \% V is enough to suppress totally the SDW state in
Cr. The decrease of the Cr magnetic moment in dilute CrV alloys
has recently been confirmed by M\"{o}ssbauer spectroscopy studies
with $^{119}$Sn probe atoms \cite{dubiel96}.

From the above mentioned facts it is reasonable to expect a strong
suppression  of the Cr magnetic moment and the SDW near the Cr/V
interface.  Recent theoretical calculations of the Cr/V interface
magnetic structure performed by different groups
\cite{{boussendel98}, {buhlmayer00}, {hamad01}, {kellou03,
herper03}} confirm this notion. In addition, the calculations
predict that the boundary V monolayer should gain an induced
magnetic moment, similar to the Fe/V interface, where theoretical
and experimental work established a small induced V moment,
polarized opposite to the Fe moments \cite{Schwickert,
Scherz,Scherz03,Labergerie, Uzdin}. The interface effects
discussed so far are expected to be of short range and should damp
out quickly at a distance of several monolayers from the
interface. In Cr/V multilayers the damping effect should manifest
itself by locating the node of the SDW in the vicinity of the Cr/V
interfaces \cite{hirai02}. It should be noted, however, that all
theoretical investigations assume that the interface is perfectly
sharp and that the propagation of the SDW is in the direction
normal to the interface. The propagation direction of the SDW is
not a result of the various ab-initio calculations, but has been
incorporated artificially.

Mibu et al. \cite{{almokhtar00}, {mibu01}} have recently provided
experimental support for the outlined theoretical picture in a
series of elegant experiments with M\"{o}ssbauer spectroscopy by
inserting the $^{119}$Sn probe layers in Cr/V multilayers at
different distances from the interfaces. They have given direct
experimental evidence for a reduction of the Cr magnetic moment
near the interface region. Cr was found  nonmagnetic at distances
up to 20 \AA ~away from the interface, while at larger distances
of about 40 \AA  ~the Cr magnetic moment increases again and
reaches values comparable to the bulk moment.

So far we have discussed the local and short range effects which
occur close to the Cr/V interface. Next we draw our attention to
the more global properties. For instance, it would be highly
interesting to investigate whether the Cr/V proximity effect
changes the N\'{e}el temperature and the SDW parameters, including
period, polarization, and propagation direction. Some theoretical
work points to these more global effects. According to Ref.
\cite{{dubiel02},{cieslak03}}, surface effects in isolated Cr
films may extend up to several thousands of {\AA}ngstroms. In a
similar fashion the Cr/V interface may perturb the Cr SDW
magnetism more extensively than assumed so far. This has to be
verified by experiments.

In the present paper we address the above questions concerning the
global perturbation of the Cr SDW magnetism by Cr/V interfaces and
provide a systematic study of the SDW state in 2000 {\AA} thick
Cr(001) films grown on a 14 {\AA} thick V(001) buffer layer. The
Cr thickness is chosen to be large enough to ignore any local
interface effects and to concentrate only on long-range effects
arising from the Cr/V interface. The amount of V atoms in the V
layer is comparable with a CrV alloy of corresponding V
concentration. From neutron and synchrotron scattering experiments
we establish the magnetic phase diagram of the system and reveal
the Cr/V proximity effect on the SDW state in comparison to other
Cr thin film systems, bulk Cr, and CrV alloys.

\section{\label{sec:level1}Growth and sample characterization}
The sample was grown with a UHV magnetron sputtering system on a
20x20 mm$^{2}$ MgO(001) substrate. The deposition started with a V
buffer layer of 14 {\AA} thickness and continued with a 2000 {\AA}
thick Cr film. No protection layer was deposited, so a thin oxide
layer is formed at the top of the Cr film \cite{stierle97}. During
the deposition the substrate temperature was kept at 200$^\circ$C.
After the preparation the sample was cut into several parts to be
used in different experiments. In particular, we used a 15x20
mm$^{2}$ sample for neutron measurements and a 5x5 mm$^{2}$ sample
for x-ray experiments.

The structural information on the sample was obtained with x-ray
diffraction studies done at the W1.1 and D3 beamlines of the
HASYLAB, Germany. The measurements were performed at room
temperature using a wavelength  $\lambda$ = 1.5404 {\AA}. The in-
and out-of-plane lattice parameters of the Cr film as well as the
epitaxial relationship between the film and MgO substrate were
established precisely by analyzing the scattered intensity around
the Cr (002) and (011)  fundamental peaks. The epitaxial relation
between Cr and MgO was confirmed to be Cr(001)[100]//MgO(001)[110]
that is also valid for growth of Cr directly on the MgO(001)
substrate without any buffer layer \cite{kunnen02}, which is
equivalent with the well known 45$^\circ$ epitaxial relation
observed for Fe(001) on MgO(001). The Cr lattice was found to have
a slight tetragonal distortion determined by the difference
between the out-of-plane lattice parameter
$a^{\perp}=2.8865\pm0.0005$ {\AA} and the in-plane lattice
parameter $a^{\parallel}= 2.879{\pm}0.001$ {\AA}.

\begin{figure}[here]
\includegraphics[width=8cm]{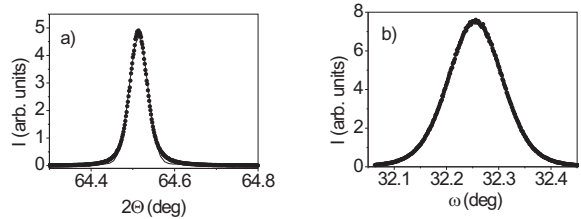}
 \caption{\label{fig1}
 a) Radial x-ray scan through the
Cr(002) reflection. The solid line is a fit to the data points
with a single Gaussian line shape. The width of the peak depends
on the Cr film thickness and reflects a 100\% structural coherence
of the film;
 b) Transverse x-ray scan
through the Cr(002) reflection. The solid line is a fit to the
data points with a single Gaussian line shape with a FWHM of
0.10$^\circ$. }
\end{figure}

Next we discuss the structural correlation length and the film
mosaicity. In Fig.~\ref{fig1}a is shown the radial x-ray scan
through the Cr(002) reflection. The solid line represents a fit to
the data points by using a single Gaussian line shape. The full
width at half maximum (FWHM) is estimated to be $\Delta(2\theta)$
= 0.044$^\circ$.
According to the Scherrer equation \cite{warren90}, this FWHM
value yields an out-of-plane structural coherence length of 2000
{\AA}, which corresponds exactly to the total Cr film thickness.
In Fig.~\ref{fig1}b is presented the transverse x-ray scan through
the Cr(002) reflection. The solid line represents a fit to the
data points by using a single Gaussian line shape. The FWHM is an
estimate of the mosaic spread in the Cr film, which is about
0.10$^\circ$. This is 5 times less than reported earlier for Cr
growth on MgO(001) \cite{kunnen02} and about 2 times less than Cr
growth on a Nb(001) buffer layers \cite{bodeker99}. Thus, the
initial growth of a thin V buffer layer turns out to improve  the
Cr intralayer structure significantly.

\section{\label{sec:level1}Scattering experiments}
When dealing with Cr thin film systems, one should distinguish
between in-plane and out-of-plane SDW propagation directions and
polarizations  that are nonequivalent due to the broken symmetry
at the interfaces. The Cr magnetic structure can be explored by
elastic neutron scattering providing complete information on the
SDW state. The magnetic moment modulation produces corresponding
satellite reflections around forbidden bcc Bragg reflections,
which can be detected with neutron scattering.

The SDW is accompanied by a charge density wave (CDW) and a strain
wave (SW), corresponding to periodic modulations of the charge
density and the lattice spacing, respectively. The modulation
period of the SW and CDW is half the period of the SDW. The SW can
be investigated via x-ray scattering, yielding information about
the SDW period and propagation direction \cite{{tsunoda74},
{gibbs88}, {hill95}}. Although SW investigations do not reveal the
SDW polarization, they have the advantage of a higher reciprocal
space resolution, which is important for a precise determination
of the SDW period. The application of neutron and X-ray scattering
methods to the determination of the SDW parameters has been
reviewed in some detail in a number of papers. We will not repeat
this here and refer the interested reader to published papers and
reviews for further information \cite{{fawcett88}, bodeker99,
zabelNATO, {zabel99}}.

\subsection{\label{sec:level2}Synchrotron scattering measurements}
Synchrotron scattering experiments were performed at the ID20
magnetic scattering undulator beamline of the ESRF (Grenoble,
France) \cite{{stunault98}, {mannix01}}. The incident photon beam
delivered by two phase undulators was 99.8 \% linearly
$\sigma$-polarized in the sample plane. The primary slits in the
optical system were set to utilize the first harmonic of the
undulators. The incident beam energy was selected to be slightly
below the Cr absorption edge of 5.989 keV by using a Si(111)
double crystal monochromator. The polarization of the scattered
beam was analyzed by using a pyrolytic graphite PG(004) analyzer
to select pure  $\sigma\sigma$ scattered intensity, also providing
a drastic reduction of the background signal. The measurements
were taken at temperatures between 15 and 300 K by using a displex
cryostat equipped with Be windows.

\begin{figure*}
\includegraphics[width=15cm]{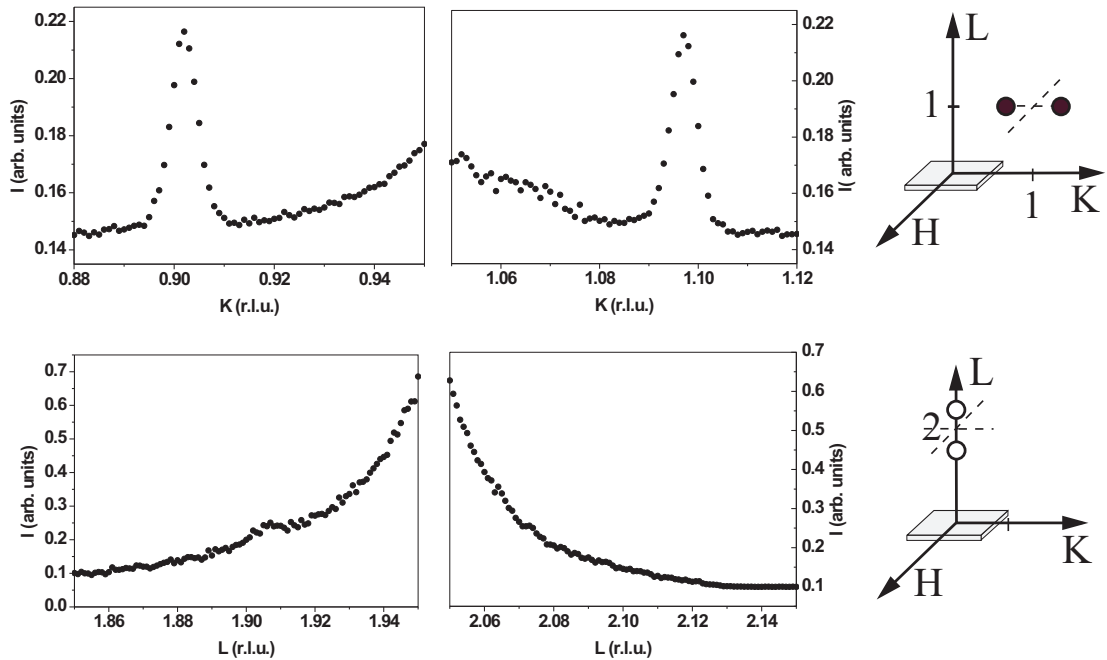}
 \caption{\label{fig2} Synchrotron radiation
scans (a) along the K direction around the Cr(011) peak and (b)
along the L direction around the Cr(002) reflection taken at 15K
without polarization analysis. The fundamental (002) and (011) Cr
peaks are removed from the figures since their intensity is many
orders of magnitude higher than the intensity of the satellite
reflections.}
\end{figure*}

To determine the propagation direction of the SW and SDW we first
performed two screening scans: one scan at the Cr(011) reflection
in the K direction to check possible in-plane SW propagation, and
one scan along the L direction, crossing the Cr(002) reflection to
search for out-of-plane SW. In Fig.~\ref{fig2} are shown the above
scans measured at 15K without polarization analysis. The
fundamental (002) and (011) Cr peaks are removed from the figures
for clarity as their intensity is many orders of magnitude higher
than the intensity of the satellite reflections. In the panel on
the right side the scan directions are indicated. The circles
refer to satellite reflections allowed by the selection rules but
not detected, whereas closed circles refer to allowed and detected
satellites. From the recorded scans it is evident that in our case
the strain wave propagates entirely in the film plane, whereas the
out-of-plane wave is completely suppressed. In subsequent
experiments we have analyzed in more detail the in-plane SW
component.
\begin{figure*}
\includegraphics[width=15cm, height=6cm]{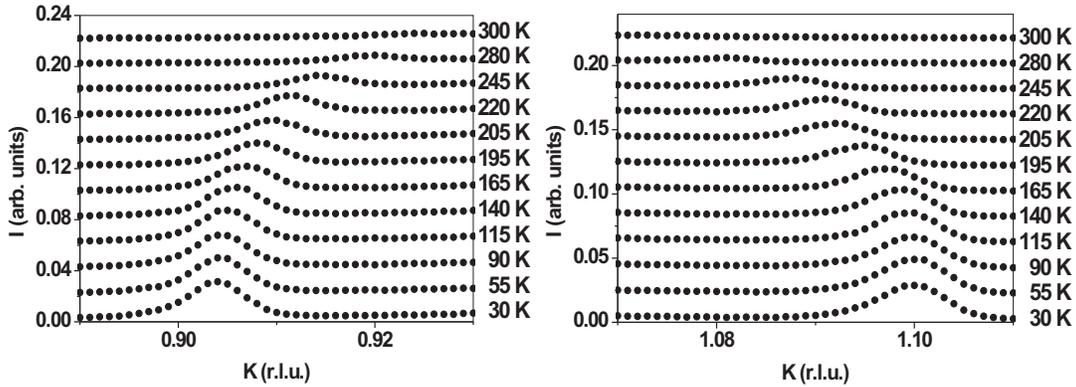}
 \caption{\label{fig3} Synchrotron measurement
of the temperature dependence of the scattered intensity along K
direction around the Cr(011) reflection taken with polarization
analysis of the scattered beam. The fundamental  Cr(011) peak is
removed from the figure, the curves and the individual scans are
shifted vertically by a constant factor for clarity.}
\end{figure*}
\begin{figure*}
\includegraphics[width=15cm]{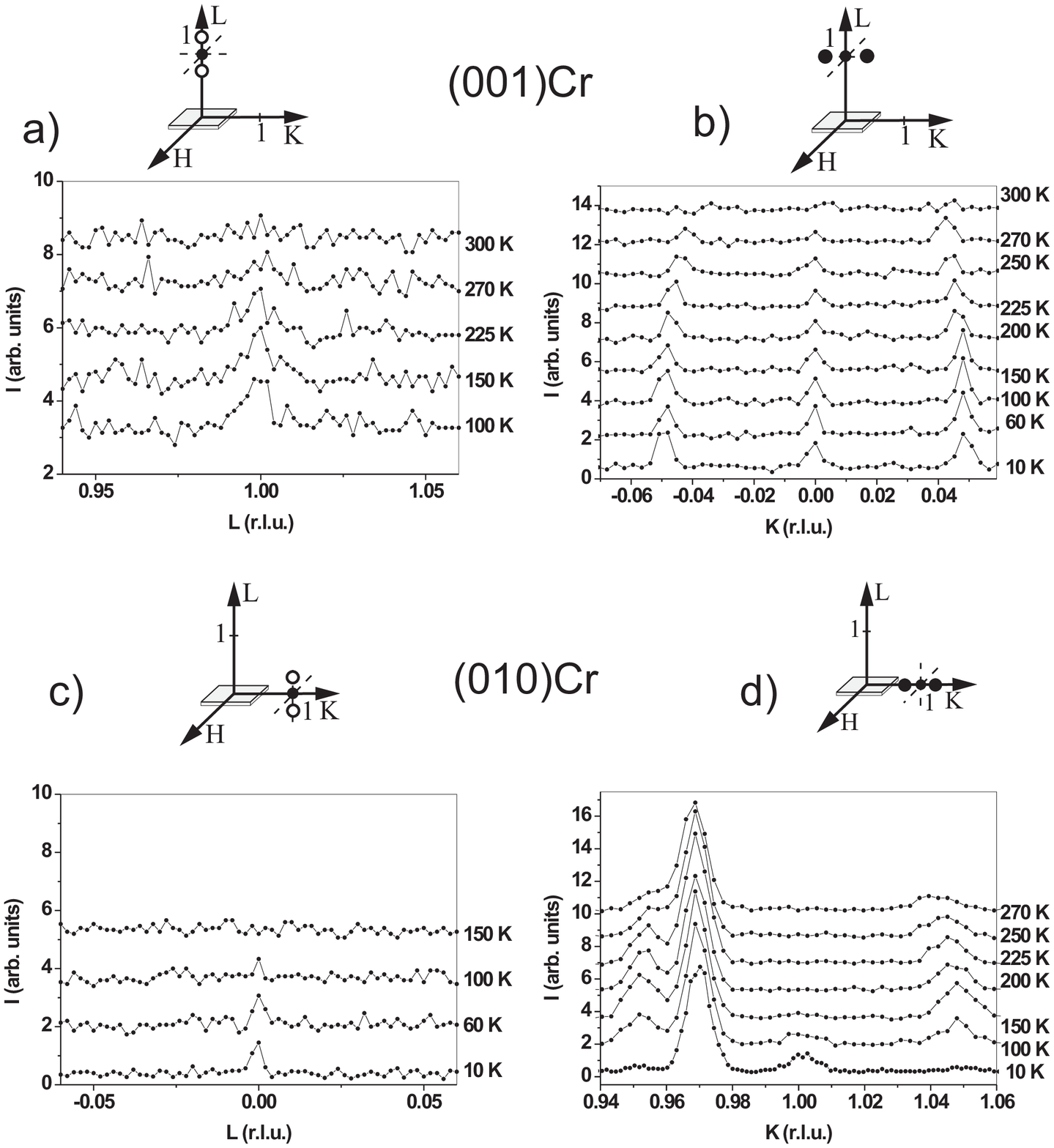}
 \caption{\label{fig4} Neutron scattering
scans taken at the Cr(010) and the Cr(001) positions to explore
the SDW polarization. The scans recorded at different temperatures
are depicted with an offset in the vertical direction for clarity.
The strong temperature-independent peak at K$\sim$ 0.97 r.l.u. in
Fig.~\ref{fig4}d is due to the MgO(022) reflection for the
$\lambda/2$ harmonics.}
\end{figure*}
In Fig.~\ref{fig3} we show the satellites position and intensity
measured for different temperatures in the range between 30 and
300 K. As evident from the picture, the satellites positions move
smoothly towards the (011) peak as temperature increases, while
the intensity decreases continuously with temperature. It is
important to note that the FWHM of the satellite peaks, about
0.007 in r.l.u., is essentially independent of temperature over
the entire temperature region investigated. At 300 K the satellite
peak intensity has vanished, which can be taken as the N\'{e}el
temperature of our system. The information on the SW behavior
inferred from the synchrotron scattering data will be discussed
further below together with the neutron scattering results.

\subsection{\label{sec:level2}Neutron scattering measurements}
The neutron scattering experiments were performed on the
triple-axis spectrometer UNIDAS installed at the FRJ-2 research
reactor (Forschungszentrum J\"{u}lich, Germany). In the experiment
we used the (002) reflection of a highly oriented PG double
crystal monochromator to select a neutron beam wavelength of
$\lambda=2.351$ \AA. Another PG analyzer and a system of special
slits and collimators were used to improve the instrumental
resolution by reducing the neutron beam divergence and the
background radiation. For these experiments the spectrometer was
operated as a diffractometer by fixing the analyzer crystal to
zero energy transfer. In this mode the background radiation is
effectively filtered out. The contamination of the $\lambda/2$
radiation was removed by placing a PG transmission filter in the
incident neutron beam. The measurements were taken at temperatures
between 30 and 300 K by using a displex cryostat with Al windows.

In Fig.~\ref{fig4} are presented neutron diffraction scans
performed in the vicinity of the Cr(010) and Cr(001) positions in
the reciprocal space. Above each set the scan directions are
indicated. Open circles refer to satellite reflections allowed by
the selection rules but not detected, whereas closed circles refer
to allowed and detected reflections. The above neutron results are
sufficient for a complete analysis of the SDW parameters including
their temperature dependence. We note that the neutron scattering
results are in complete agreement with the synchrotron scattering
results. As seen in Fig.~\ref{fig4}, the satellite reflections due
to the incommensurate SDWs occur only in the K direction in the
vicinity of the (001) and (010) reflections. This implies that the
SDW propagates in the film plane, as already conjectured from the
x-ray scans. The N\'{e}el temperature is estimated to be about 300
K, which agrees with the synchrotron results. The temperature
dependence of the SDW polarization is nearly identical to the one
observed in bulk Cr \cite{werner67}. At low temperatures we
observe a longitudinal SDW.  Above 100 K a spin-flip transition
occurs to an in-plane transverse SDW. However, in contrast to the
bulk behavior, the longitudinal and the transverse SDW are
bi-domain, and in both cases the Cr magnetic moments are aligned
parallel to the film plane.

In addition to the satellite reflections from ISDW phase, we also
observe the (001) reflection corresponding to CSDW. The
commensurate phase coexists with the incommensurate one at all
temperatures up to the N\'{e}el temperature. From Fig.~\ref{fig4}
we notice that the intensity of the (001) reflection decreases
smoothly with increasing temperature up to 300 K, while the (010)
peak intensity disappears above 150 K. Thus the polarization of
the CSDW is similar to the ISDW: the magnetic moments remain in
the film plane for all temperatures. Below 100 K the magnetic
moments are oriented parallel to the H direction, and above 150 K
parallel to the K direction. The transition from one to the other
polarization direction occurs continuously between 100 K to 150 K.
The integrated intensity of the CSDW reflection is about 10 \% of
that of the ICDW satellites. Both the CSDW and ISDW reflections
have the same FWHM, which is directly connected  with the in-plane
SDW correlation length.

\begin{figure}[here]
\includegraphics[width=10cm]{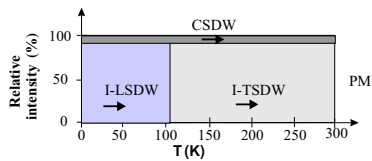}
 \caption{\label{fig5} Qualitative magnetic phase diagram for the
 spin density wave in a 2000 \AA ~thick Cr(001) film deposited on
 a MgO(001) substrate with a 15 \AA ~thick V buffer layer. The phase diagram
 is the result of combined synchrotron and neutron-scattering
 experiments, as described in the text.
}
\end{figure}

In Fig.~\ref{fig5} is depicted the temperature phase diagram for
the V/Cr system. Both the ISDW and the CSDW propagate always in
the film plane with the magnetic moments lying also in the film
plane. The spin-flip transition for both phases starts at the same
temperature of about 100 K. While the spin-flip transition for the
ISDW is more or less sharp, the spin reorientation transition for
the CSDW is spread over a wide temperature interval of about 50 K.
\section{\label{sec:level1}Discussion}

The synchrotron and neutron scattering results described in the
previous section allow us to draw a clear picture of the SDW
behavior within the Cr/V system. To work out the essential details
arising from the Cr/V proximity effect, we will compare the
results obtained with those known from reference systems: bulk Cr,
dilute CrV alloy and single Cr films on other paramagnetic or
ionic substrates.

First we discuss possible effects of the vanadium buffer layer on
the SDW polarization and propagation direction. It is generally
recognized now that the SDW propagation direction in thin Cr films
is governed by strain effects from substrate and buffer layer,
whereas the SDW polarization is mainly due to proximity effects
from neighboring layers.  In case of a 4000 \AA ~thick  Cr(001)
film on a Al$_{2}$O$_{3}$(1\={1}02) substrate with a Nb(001)
buffer layer a mixture of in-plane transverse and out-of-plane
longitudinal ISDW has been observed below the N\'{e}el
temperature, while for a 3000 \AA ~thick Cr film an out-of-plane
LSDW is observed at low temperature, which transforms to a
out-of-plane TSDW at higher temperatures
 \cite{sonntag98,sonntag95}. In all cases the Cr surfaces was
uncapped and therefore was covered with a roughly 20 \AA ~thick
Cr$_{2}$O$_{3}$ layer, which forms by natural oxidation
\cite{stierle97}. For a single 4500 \AA~ Cr film grown on the
MgO(001) substrate without any buffer layer, Kunnen et al.
\cite{kunnen02} observed a mixture of CDSW and  ISDW, which
propagate in the film plane, while the spins are oriented in the
direction normal to the film plane. Our recent data for a 2000
\AA~ Cr film at MgO(001) substrate \cite{unpublished} are in line
with the results \cite{kunnen02}. We found that the SDW propagates
in the film plane, it has longitudinal polarization at
temperatures below 100 K and transversal with out-of-plane spins
at temperatures from 100 to 270 K. At higher temperatures we
observed the in-plane CSDW with out-of-plane spins. The
propagation direction agrees with our observations, but not the
spin direction. So we conclude that the proximity effect from thin
V layer is responsible for the SDW polarization.

This unusually strong proximity effect is unexpected and it is
difficult to give its adequate explanation. Based on present
state-of-art in the field, we would only suggest that the
underlying mechanism is connected with induced magnetic
polarization at the V/Cr interface. As mentioned in the
Introduction, at V/Fe and at V/Co interfaces an induced V moment
polarized antiparallel to the ferromagnetic moments has been
established \cite{Scherz,Scherz03,Huttel03}. It is possible that a
similar polarization may also occur at the V/Cr interface. In this
case, the V magnetic moments are aligned parallel to the interface
and are exchange coupled indirectly with the Cr magnetic moments
across a non-magnetic layer at the interface \cite{almokhtar00}.
This might explain the difference in spin orientation for Cr on
plain MgO (out-of-plane) and for Cr on MgO with a thin buffer
layer (in-plane).

The actually observed SDW may  not only depend on the electronic
properties of the boundary layer, but also on the strain state and
thickness of Cr. To the best of our knowledge, there is no
theoretical prediction available as concerns the orientational
effects caused by paramagnetic boundary layers. We note, however,
that the effect of pinning the SDW nodes near the Cr/V interfaces
as predicted by Hirai \cite{hirai02} does not apply in our case,
since the SDW propagates in the layer plane.

\begin{figure}[here]
\includegraphics{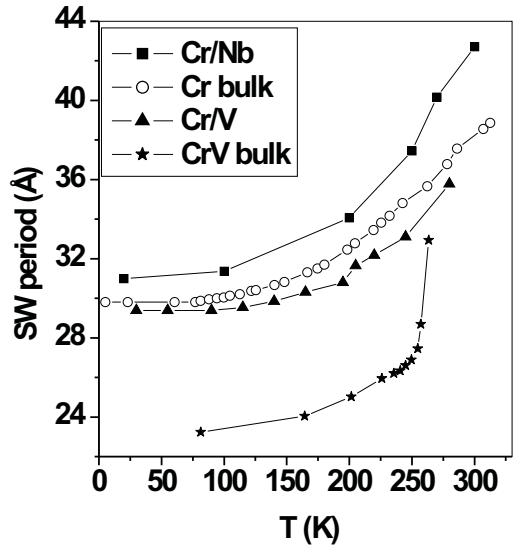}
 \caption{\label{fig6} Temperature dependence of the strain wave period
 in the Cr/V film and corresponding data for reference systems:
 2000 \AA ~Cr film on a  thick Nb buffer layer \cite{sonntag98}, bulk
 Cr \cite{hill95}, and in bulk Cr$_{0.995}$V$_{0.005}$ alloy
 \cite{noakes88}. The solid lines connecting the data points are
 guides to the  eye.
 }
\end{figure}

Next we consider proximity effects of vanadium on other SDW
parameters, in particular, on the SDW period. Since the
synchrotron data on the SW are more precise than the neutron data,
we use those to estimate the period of the SDW as a function of
temperature. The data are plotted in Fig.\ref{fig6} and compared
to the SDW period in bulk Cr \cite{hill95} and in a 2000 \AA~
thick Cr film grown on a thick Nb buffer layer \cite{sonntag98}.
Since no data are available for thin CrV films, we compare our
results with a  bulk Cr$_{0.995}$V$_{0.005}$ alloy
\cite{{noakes88}, {camargo99}, {oliveira95}}. This corresponds
roughly to the same composition as Cr(2000 \AA)/V(14 \AA) when
mixed, which would result in a 0.7 at \% alloy.
As can be seen from the figure, the temperature dependence of the
SW period in our Cr/V film is similar to that in other Cr thin
film systems and in bulk Cr, but completely different from that in
the Cr$_{0.995}$V$_{0.005}$ alloy. The value of the SW period in
the Cr/V film is smaller than in bulk Cr and in other thin Cr
films, but much larger than in the corresponding bulk CrV alloy.
To our knowledge, the SW period observed in our Cr/V system is the
smallest one observed so far in Cr-based thin films and
multilayers.

\begin{figure}[here]
\includegraphics{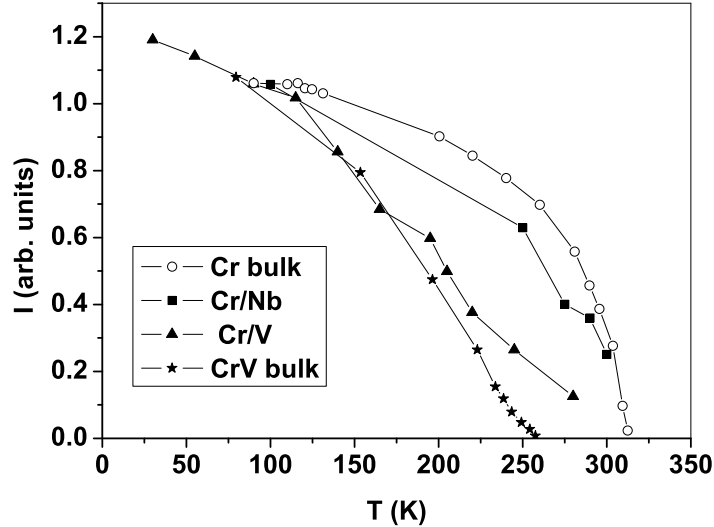}
 \caption{\label{fig7} Temperature dependence of the integrated intensity of the strain wave
 satellite reflections in the present Cr/V film and corresponding data for reference systems:
 2000 \AA ~Cr film at thick Nb buffer\cite{sonntag98}, bulk
 Cr \cite{hill95}, and integrated intensity of the spin density wave
 satellite reflections  measured with neutron scattering in bulk CrV alloy with a V
 concentration of  0.5 at. \% (from Noakes et al.\cite{noakes88}).
 }
\end{figure}

In  Fig.~\ref{fig7} we show the temperature dependence of the
integrated intensity of the SW satellite peaks for the same
systems. The integrated intensity $I(t)$ is proportional to the
square of the order parameter which, in turn, is proportional to
the Cr magnetic moment. Its temperature dependence is a
characteristic of the phase transition from the antiferromagnetic
to the paramagnetic state in Cr. It is known that the temperature
dependence $I(t)$ in bulk Cr can be described in analogy to the
BCS theory of superconductivity \cite{overhauser62}.

In the present Cr/V system we find a quasi-linear temperature
dependence of $I(t)$ at low temperatures, which resembles the
temperature dependence observed in CrV alloys \cite{noakes88}, but
which is drastically different from the $I(t)$ dependence seen in
bulk Cr and other thin films. Noakes et al. \cite{noakes88}
speculated that the unusual $I(t)$ behavior may not be an
intrinsic feature of the CrV alloy, but due to strain effects.
From our analysis it appears that the $I(t)$ dependence is an
intrinsic property of the CrV systems. At higher temperatures the
$I(t)$ for Cr/V deviates from that of the CrV alloy and stretches
towards the bulk N\'{e}el temperature. Another important
difference is the temperature dependence for the FWHM of the SW
satellite peaks in Cr/V as compared to bulk Cr. According to Hill
et al \cite{hill95}, in bulk Cr the FWHM of the SW satellite
reflection increases with temperature, implying a reduction of the
SDW correlation length with increasing temperature. In Cr/V we
find no change in the FWHM of the SW satellite peak over the
entire temperature range and consequently no change in the
correlation length.

\section{\label{sec:level1}Conclusions}
We have studied the proximity effect of vanadium on the spin
density wave magnetism in Cr/V films. The sample was grown with a
UHV magnetron sputtering system on MgO(001) substrate with a 14
\AA ~thick V(001) buffer layer.  The Cr film exhibits a very high
structural quality expressed by an out-of-plane coherence length
corresponding to the film thickness and a very small mosaicity.
The SDW properties were investigated by a combination of
synchrotron scattering experiments to probe the strain waves and
elastic neutron scattering to probe the spin density waves in the
Cr film. It was shown that the V-Cr hybridization at the interface
causes a strong and long-range influence on the SDW behavior in
the Cr film. First, we found that the V-Cr interface hybridization
changes the SDW polarization from out-of-plane to in-plane, i.e.
the transverse incommensurate SDW propagates in the film plane and
the magnetic moments are also  in the plane. Second, for the SDW
period and the order parameter we see a mixture of features which
are typical for both, bulk CrV alloys and simple Cr films. The
temperature dependence of the SDW period is similar to that in
other Cr films but different from the one in CrV alloys. On the
other hand, the value of the SDW period is smaller than in bulk Cr
and other Cr films, which is a characteristic feature of CrV
alloys. The Cr magnetic moment in Cr/V bilayers decreases
quasi-linearly at low temperatures, which is also observed in some
CrV alloys but not in bulk Cr or other Cr-based thin film systems.
The N\'{e}el temperature corresponds to the bulk value. Above the
N\'{e}el temperature for the incommensurate phase we do not
observe any commensurate SDW, which we ascribe to the high quality
of the film. In any case, the commensurate SDW plays only a minor
role in the present Cr/V film system.

From the present experiments we conclude that the effect of a very
thin V layer (14 \AA) on a thick Cr film (2000 \AA) is
surprisingly large, changing drastically the global features of
the SDW over the entire Cr film. This concerns the polarization of
the SDW, the period, and the temperature dependence of the Cr
magnetic moment at low temperatures. However, the N\'{e}el
temperature is roughly the same as in the bulk.

\begin{acknowledgments}
We would like to thank Dr. O. Seek and Dr. W. Morgenroth for help
with the beamlines operation at the W1.1 and D3 instruments of the
HASYLAB. This work has benefitted from collaborations within the
Sonderforschungsbereich 491 'Magnetische Heteroschichten: Struktur
und elektronischer Transport' funded by the Deutsche
Forschungsgemeinschaft and from international collaborations
supported by INTAS under project No. 01-0386. E.K. acknowledges
support from RFBR.
\end{acknowledgments}

\newpage 

\end{document}